\documentclass[a4paper]{jpconf}
\usepackage{graphicx}

\begin{document}

\title{Instability in a magnetised collisional plasma driven by a heat flow or a current}

\author{AR Bell$^{1,2}$,  RJ Kingham$^3$, HC Watkins$^3$, JH Matthews$^4$}

\address{
$^1$ Clarendon Laboratory, University of Oxford, Parks Road, Oxford OX1 3PU, UK
\newline
$^2$ Central Laser Facility, Rutherford Appleton Laboratory, Chilton, OX11 0QX, UK
\newline
$^3$ Blackett Laboratory, Imperial College, London SWZ 2BZ, UK
\newline
$^{4}$ Institute of Astronomy,  Madingley Road, Cambridge CB3 0HA, UK
}

\ead{Tony.Bell@physics.ox.ac.uk}

\begin{abstract}
We solve the linearised Vlasov-Fokker-Planck (VFP) equation to show that heat flow or an electrical current in a magnetized collisional plasma 
is unstable to the growth of a circularly polarised transverse perturbation to a zeroth order uniform magnetic field.  
The Braginskii (1965) transport equations exhibit the same instability in the appropriate limit.
This is relevant to laser-produced plasmas, inertial fusion energy (IFE) and to dense cold interstellar plasmas.
\end{abstract}

\section {Introduction}

In plasmas,  particle momentum distributions are rarely Maxwellian and isotropic.  
They frequently carry a large energy flux or a component of energetic particles.  
This is especially so where there is a large localised deposition of energy.
A dimensionless measure of an energy flux is its ratio to the so-called free-streaming heat flow, $Q_{f} = n_em_ev_t^3$, 
where $n_e$ is the electron density and $v_t=\sqrt{eT/m_e}$ is the  electron thermal velocity 
if the plasma can be characterised by a 
 temperature $T$ in eV. 
In ablating systems, the heat flow is naturally of the order of 10\% of the free-streaming heat flow. 
This is particularly a feature of laser-irradiated dense plasmas that constitute the main interest of this paper (eg Craxton et al 2015).
A number of plasma instabilities are known to be associated with strong heat flow in laser-plasmas.
These include the Tidman \& Shanny instability (Tidman \& Shanny 1974, Sherlock \& Bissell 2020), the Weibel instability (1959), the electrothermal instability (Haines 1981),
the ion-acoustic instability that generates an anomalous resistivity (Forslund 1970, Rozmus et al 2018),
and the field-compressing magnetothermal instability of Bissell et al (2010).
Apart from the field-compressing magnetothermal instability
these instabilities do not require a pre-existing zeroth order magnetic field.
The Tidman-Shanny instability is driven by the Biermann battery and relies upon the presence of the density gradient found in ablating plasmas. 
The Weibel instability is essentially collisionless and grows on the scale of a skin depth $c/\omega _{pe}$. 
In contrast, the instability considered in this paper requires the existence of a zeroth order magnetic field which is parallel to the heat flow.
The field-compressing magnetothermal instability combines Nernst advection and
Righi-Leduc heat flow to amplify a pre-existing magnetic field that is perpendicular to the heat flow.

Plasma instabilities driven by heat flow are also well-known in space plasmas.
These have been studied extensively over many decades
(Forslund 1970, Gary et al 1975, Gary 1985)
where they are known as `Heat Flow Instabilities'.
Many different instabilities have been identified 
(Schwartz 1980, Davidson 1983), and a recent discussion
can be found in Lee et al (2019).
Although collisions may be important in the formation of the
heat flow in solar wind plasmas, they are too weak to affect their instability on small scales.
Our analysis includes collisions, but contains the collisionless case as
a limit.
The dispersion relation derived in section 4 below gives the growth rate for a range of different unstable modes
 depending on the values of various dimensionless parameters.

Electron collisions and Larmor gyration
are essential ingredients of the instability focussed on here.
We also include the relatively slow ion response and find that it is important
in the long wavelength limit where the instability transitions into the non-resonant instability that is known to drive magnetic field amplification during
cosmic ray (CR) acceleration (Bell, 2004).
We find that magneto-collisional  instabilities are present (i) in relatively relaxed plasmas in which the energy flux is carried
by electrons in the high velocity tail of a Maxwellian distribution and (ii) in plasmas far from equilibrium in which 
an independent population of high velocity charged particles, such as cosmic rays or laser-produced energetic protons or electrons, pass through  a thermal plasma.

We consider the case in which a strong heat flow or current of energetic particles propagates parallel to a zeroth order magnetic field.
This is a common configuration since
collisionless electrons in a uniform plasma are mobile along magnetic fields lines, but cannot pass easily across a uniform magnetic field.
We add a small perturbation to the magnetic field and show that it grows exponentially.

`Cross-field' transport perpendicular to the local magnetic field can be enabled by collisions, electric field, curvature in magnetic field lines, gravitational fields, 
or hydrodynamic acceleration of the bulk plasma (equivalent to the effect of gravity).
These processes have different dependences on electron velocity.
For example, the strong velocity dependence of the Coulomb collision rate means that collisions 
can dominate at low velocities while Larmor gyration dominates at high velocities.
We find that the competition between collisions and Larmor gyration produces instability.
This `magneto-collisional instability' is strongest when $\Omega_e \tau$ is of the order of one
where $\Omega _e$ is the electron Larmor frequency and $\tau$ is the collision time of a thermal electron.
Under these conditions, an electric field is generated to maintain electrical neutrality and satisfy $\nabla \times {\bf B} = \mu _0 {\bf j}$.
The curl of this electric field causes unstable growth of the perturbed magnetic field.

\begin{figure}
\includegraphics[angle=0,width=8cm]{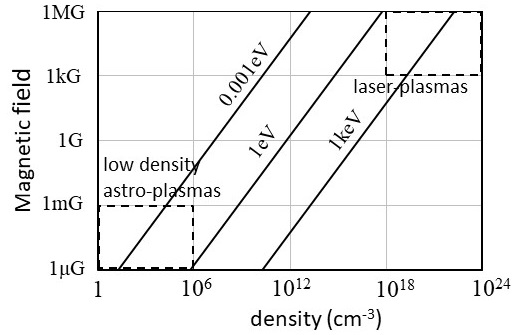}
\centering
\caption{
The lines give the temperature at which $\Omega _e \tau =1$ for a range of densities and magnetic fields
with $Z \log \Lambda=10$.
}
\label{fig:figure1}
\end{figure}

The conditions for $\Omega _e \tau \sim 1$ are plotted in figure 1.
As shown in figure 1, the conditions for strong growth occur naturally in laser-produced plasmas.
In low density astrophysical plasmas relevant to CR acceleration, the collisionality is usually low, $\Omega_e \tau \gg 1$,
but the magneto-collisional instability may be strong where CR propagate into a dense cold cloud,
or where a dense cold cloud is ablated by thermal conduction with a heat flow close to free-streaming.
The magneto-collisional instability may also be relevant in low density weakly collisionless plasmas where the growth time is of the order of the
collision time which, despite being long compared to the Larmor gyration time, may be short compared to the timescale for hydrodynamic evolution.

Since both collisions and Larmor gyration are important, the derivation of the general dispersion relation requires solution of the Vlasov-Fokker-Planck (VFP) equation
in which advection,  collisions, the effect of magnetic and electric field, and high order anisotropies are modelled.

This paper is structured as follows.
The central core of the paper is contained in sections 3 to 6.
Section 3 sets out the basic equations consisting of the Maxwell equations for the electromagnetic fields, the VFP equation for the electron response, 
and the cold ion fluid equations.  Section 4 derives the dispersion relation.
Section 5 solves the dispersion relation for a heat flow parallel to a magnetic field.
Section 6 solves the dispersion relation for a thermal plasma carrying a return current that balances the electric current carried
by streaming energetic charged particles.

Before embarking on the main general derivation in section 3, we insert in section 2 a demonstration of instability in a simple special case.
Our aim in section 2 is to orientate the reader to the basic physics underlying the mathematics in 
the following sections.

Following on from the VFP derivations in sections 3 to 6,
we show in section 7 that the long wavelength limit of the same instability can be derived from the Braginskii transport equations.
Braginskii's analysis starts from the Boltzmann equation which is equivalent to the VFP equation.
Braginskii solves the linearised equations  in the 
limit of spatial scalelengths much larger than  the electron  Larmor radius and the electron mean free path.
We find that the Braginskii and VFP growth rates are mutually consistent
within Braginskii's scheme of approximation.


\section{Setting the scene}

The dispersion relation derived in later sections of this paper contains a number of unstable modes. 
In this section we illustrate the common underlying principle by a simple demonstration of instability in one specific configuration.

Consider a laser-plasma experiment in which a high intensity laser pulse is directed at the front surface of a thin solid foil.
The interaction produces a stream of energetic protons that emerges from the rear of the foil with an efficiency that can reach 
$\sim 10 \%$ (eg Snavely et al 2000).  
The experimental parameters can be chosen such that the protons have energies in the range 
100keV-10MeV.  
The streaming protons are then passed through a second `target' plasma.
Because of their high momentum, the protons form a directed rigid beam that is only weakly deflected
by collisions and electromagnetic fields.
Neutrality and magnetic induction require the electrons in the target plasma to carry a return current
that cancels the electrical current carried by the energetic protons.
The small deflection of the energetic protons as they pass through the target plasma is often used for proton radiography
to diagnose a target plasma (eg Mackinnon et al 2004, Tzeferacos et al 2018).
The target plasma might contain magnetic fields generated by a variety of processes, for example by
the Biermann battery.
Consider the case in which the target plasma is threaded by a magnetic field ${\bf B}_0$ that is aligned
in the direction of the proton beam.  
Such a  magnetic field may arise naturally in the experiment with a magnitude in the range 10kG - 1MG.
Or alternatively, in an experiment designed to investigate the response of the target,
an external magnetic field could be imposed with a magnitude that can exceed 1MG (eg Santos et al, 2018).
We now show that this configuration is unstable to the growth of helical perturbations in the magnetic field.

\begin{figure}
\includegraphics[angle=0,width=7cm]{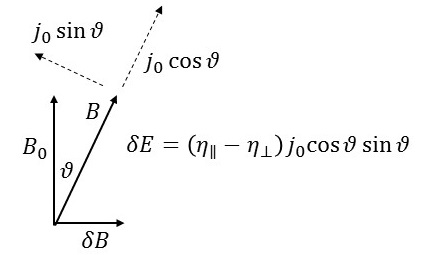}
\centering
\caption{
Components of the current (dotted lines) when a perturbed magnetic field $\delta {\bf B}$ is added perpendicular to the zeroth order field ${\bf B}_0$.
}
\label{fig:figure2}
\end{figure}

Figure 2 illustrates the role of a perturbed transverse magnetic field  $\delta {\bf B}$ in generating an electric field with a curl that
causes the perturbed magnetic field to grow exponentially.
The streaming protons, passing with negligible deflection through the target plasma, draw a return current $j_0$ in the opposite direction to
the proton flux and parallel to the zeroth order magnetic field.
Since the target plasma is collisional, an electric field is needed to draw the return current.
The resistivity of the plasma is $\eta _\parallel$ parallel to the local magnetic field, and $\eta _\perp$
perpendicular to the local field.
The perturbed magnetic field $\delta {\bf B}$ is transverse to the zeroth order field ${\bf B}_0$, so the return current
has a component perpendicular to the local magnetic field and a component aligned with the local magnetic field.
As shown in figure 2, the electric field needed to draw the return current has a component $ \eta _\parallel j_0 \cos \theta $ parallel to the local magnetic field,
and a component $ \eta _\perp j_0 \sin \theta $ perpendicular to the local field.
The component of the electric field perpendicular to ${\bf B}_0$ in the plane of figure 2 is
$\delta E = (\eta _ \perp - \eta _\parallel) j_0 \cos \theta \sin \theta $  perpendicular to the ${\bf B}_0$.
In the linear limit of $|\delta {\bf B}| \ll |{\bf B}_0|$,
$\delta {\bf E}=(\eta _\parallel - \eta _\perp) j_0 \delta {\bf B} /B_0$.
The curl of $\delta {\bf E}$ increases the perturbed magnetic field $\delta {\bf  B}$ and causes exponential unstable growth of the perturbation.
The equation for the evolution of the perturbed field is
$$
\frac {\partial ( \delta {\bf B})}{\partial t}+ (  {\bf v}_j . \nabla ) ( \delta {\bf B}) 
=(\eta _\perp - \eta _\parallel ) \frac {j_0}{B_0} \nabla \times (\delta {\bf B})
\eqno{(1)}
$$
where ${\bf v}_j$ is the Nernst-like advection velocity parallel to ${\bf j}_0$ arising from the component of the electric field out of the plane of figure 2.
The resulting dispersion relation for wavenumber $k$ and frequency $\omega$ is
$$
\omega -kv_j=\pm \frac {i k v_0}{\Omega _e \tau} \frac { \Delta \eta }{\eta _ 0}
\eqno{(2)}
$$
where $\Delta \eta = \eta _\perp - \eta _\parallel$, 
$v_0 = j_0/n_e e$ is the mean velocity of electrons as they carry the current $j_0$,
and 
$\eta _0 =m_e/{n_e e ^2 \tau}$ is a fiducial resistivity.
The $\pm$ sign depends on the sense of rotation of the helical instability. 
One sense of rotation is stable and the other is unstable.

Our purpose in the derivation of equation 2 is to illustrate the physics without recourse to  complicated mathematics.
Detailed calculations are presented in the rest of the paper
where we 
show that, in the limit appropriate to this section, the maximum growth rate 
occurs when $\Omega _e \tau \sim 0.1 - 1$ (see figure 6).
For $Z=\infty$,
Braginskii (1965) gives  $\Delta \eta /\eta _0=0.038 $
($\eta _\parallel=0.294 \eta _0$, $\eta _\perp  =0.332  \eta _0$) for  $\Omega _e \tau =0.1$
and $\Delta \eta /\eta _0=0.16$
($\eta _\parallel=0.294 \eta _0$, $\eta _\perp =0.454 \eta _0$)
for   $\Omega _e \tau =1$.
The maximum growth rate is correspondingly
$$
\gamma _ {\max} \sim 0.3 kv_0\ .
\eqno{(3)}
$$

As an example, a numerical estimate of the growth rate for characteristic experimental parameters can be obtained as follows. 
The electron return current must be equal and opposite to the proton forward current if neutrality is to be maintained.
If the protons of energy $T_p $ in eV carry an energy flux $Q_p$, then $j_0=Q_p/T_p$.
As an order of magnitude estimate,
the maximum growth time $\gamma ^{-1}$ is $\sim 30{\rm psec}$ if $B_0=100{\rm kG}$, $Q_p=10^{16}{\rm Wcm}^{-2}$, $T_p=100{\rm keV}$, 
$T_e=1 {\rm keV}$,  $n_e=10^{22}{\rm cm}^{-3}$, and $k$ is equal to the inverse Larmor radius of the thermal electrons.

The essential physics of the instability arises from  the difference in electron mobility across and along a magnetic field.
An electric field is needed to ensure that the electron  return current is exactly anti-parallel to  the forward proton current.
The curl of the electric field  generates the magnetic field.

As shown below, a related form of the instability occurs during thermal electron transport.  The heat-carrying electrons on
the tail of the Maxwellian  distribution play the role of 
the streaming protons, while the thermal majority of the electrons  carry a return current.


\section{Instabilities derived from the VFP equation}
Beginning from the electron VFP equation for a plasma that is magnetised and collisional we derive a dispersion relation when the plasma carries a heat flow or an electric current.
We include the self-consistent motion of ions, but assume that the ions are cold.
The derivation can be applied to an arbitrary zeroth-order electron distribution function $F(v)$, but here we limit our 
discussion to zeroth order distribution functions that can be expressed as the first two terms
in a Chapman-Enskog expansion in degrees of anisotropy:
$$
F({\bf v})=F_0(|{\bf v}|) +({\bf v}/|{\bf v}|)F_1(|{\bf v}|) \ .
\eqno{(4)}
$$
Note that the subscripts `0' and  `1'  in $F_0$ and $F_1$ relate to the order of anisotropy, and not to unperturbed and perturbed (zeroth order and first order) quantities.
$F_0$ is taken to be Maxwellian.
$F_1$ is determined by solution of the linearised VFP equation.
Equations 20 and 21 give its form for a plasma carrying a heat flow and a current respectively.
In contrast, the perturbed part $f({\bf v})$   of the electron distribution function is allowed arbitrary anisotropy and expressed as an expansion in spherical harmonics,
$$
f(x,{\bf v},t)=\sum _{n=0}^{n_{max}} \sum _{m=-n}^{m=n}f_n^m (x,|{\bf v}|,t)P_n^m(\cos \theta)e^{im\phi}
\eqno{(5)}
$$
where $\cos\theta = v_x/|{\bf v}|$,
$\sin \theta \cos \phi =v_y/|{\bf v}|$, $\sin \theta \sin \phi =v_z/|{\bf v}|$,
$f_n^{-m}=(f_n^m)^*$, and $n_{max}$ is sufficiently large to accommodate all the relevant physics.
In some of the calculations presented here $n_{max}$ has to exceed 100.
We assume the Lorentz (high $Z$) approximation for electron collisions in which the dominant collision process is angular scattering of electrons by ions in the 
local rest frame of the ions.

The derivation proceeds by linearising the VFP equation for the $m=1$ components of the perturbed electron distribution function, 
the equations of motion for cold ions, and the Maxwell equations for electromagnetism.  
The zeroth order heat flow or the electric current (carried in each case by $F_1$) is assumed to be aligned with the zeroth order magnetic field in the $x$ direction along the unit vector $\hat{{\bf x}}$.
The perturbation is taken to be circularly polarised.
The perturbed vectors are all transverse and lie in the $(y,z)$ plane.
The equations for the  ion velocity and electromagnetic fields are
$$
n_e m_A \frac {\partial {\bf u}}{\partial t}=n_e e {\bf E}+n_e e {\bf u}\times {\bf B}+ {\bf R}
\hskip 1.2  cm
\frac {\partial {\bf B}}{\partial t}= - \nabla \times {\bf E}
\hskip 1.2 cm 
\nabla \times {\bf B}= \mu _0 {\bf J}+\mu _0 n_e e {\bf u}
\eqno{(6)}
$$
where $m_A=(A/Z)m_p=\rho/n_e$,
and the resistive frictional collisional force exerted by electrons on ions is
$$
{\bf R}=-  \int
 \left ( \frac {\partial f}{\partial t} \right ) _c
m_e {\bf v}
d^3 {\bf v}
\eqno{(7)}
$$
where 
$  \left ( {\partial f}/{\partial t} \right ) _c$
is the electron collision term.
The electric current carried by the electrons (with charge $-e$) in the lab frame, is
$$
{\bf J}= - \int e{\bf v} f  d^3 {\bf v} \ .
\eqno{(8)}
$$
The spatial differential operator $\nabla $ and the zeroth order magnetic and electric fields are  in the $x$ direction.

When expressed in terms of spherical harmonics (Bell et al 2006), the VFP equations for the $m=1$ components are ($n=1,2,3...$) 
$$
\frac {\partial f_n^1}{\partial t}=
\frac {eE_x}{m_e}\left [\frac {n-1}{2n-1} G_{n-1}^1+\frac {n+2}{2n+3} H_{n+1}^1 \right]
+ \frac {e(E_y-iE_z)}{2 m_e} \left [ \frac {G_{n-1}^0}{2n-1} - \frac {H_{n+1}^0}{2n+3} \right ] 
\hskip 8 cm
$$
$$
\hskip 3 cm 
- \frac {n-1}{2n-1} v \frac {\partial f_{n-1}^1}{\partial x}
- \frac {n+2}{2n+3} v \frac {\partial f_{n+1}^1}{\partial x}
-\frac {ieB_x}{m_e} f_n^1
+\frac {ie(B_y - i B_z)}{2 m_e}f_n^0
+ \left ( \frac {\partial f}{\partial t} \right ) _{c,n}^1
\eqno{(9)}
$$
where
$$
G_n^m=v^n \frac {\partial (v^{-n} f_n^m) }{\partial v}
\hskip 1.5 cm
H_n^m=v^{-(n+1)} \frac {\partial (v^{n+1} f_n^m) }{\partial v}.
\eqno{(10)}
$$
and  $ ( {\partial f}/{\partial t} ) _{c,n}^1 $ represents the effect of collisions:
$$
 \left ( \frac {\partial f}{\partial t} \right ) _{c,n}^1
=- \frac {1}{2} n(n+1) \nu (v)  f_n^1
- \frac {\delta _{n1}}{2} u \nu (v)   \frac {\partial F_0}{\partial v} 
\hskip .5 cm {\rm where} \hskip .5 cm
\nu (v)= \frac {3.76}{ \tau} \left ( \frac { m_e v^2}{ eT} \right )^{-3/2}
\eqno{(11)}
$$
and $\tau= \tau_{NRL}/Z$, where $\tau _{NRL}=3.44\times 10^5 (\log \Lambda )^{-1} (n_e/{\rm cm}^{-3})^{-1} (T/{\rm eV})^{3/2}$ is the collision frequency in the NRL plasma formulary
and $\log \Lambda $ is the Coulomb logarithm.
All components of the perturbed electron distribution for $m\ge 2$ are multiples of perturbed quantities and can be neglected as small in the linear analysis.

In the above,
$f_n^1$, $E_y$, $E_z$, $B_y$, $B_z$, $G_n^1$ \& $H_n^1$
are all first order perturbed quantities, and
$f_n^0$, $E_x$, $B_x$, $G_n^0$ \& $H_n^0$
are all zeroth order.
In order to distinguish the zeroth order quantities,
we label them as $f_n^0=F_n$, $E_x=E_0$ and $B_x=B_0$.

These equations result in a dispersion relation for a range of unstable modes with differing underlying physics.
Some of these are already known, especially in the collisionless limit.
Other unstable modes  are, to our knowledge, previously unknown.

In the derivation, we will assume that the frequency $\omega$ is small compared with the maximum of the electron Larmor frequency $\Omega _e$,
the electron collision frequency $1/\tau$ and the rate $kv_t$ at which electrons transit one wavelength $2 \pi /k$ of the mode.
This allows us to neglect $\partial f_n^1/\partial t$ from the VFP equation.  It removes the faster growing modes, such as velocity-resonant 
inverse Landau damping, from the dispersion relation, making the dispersion relation more easily expressed in a closed form.

We also neglect the term proportional to $E_x$ ($=E_0$) on the right hand side of equation 9.
This can be justified on either of two counts:
(i) the local approximation that $k^{-1}e E_0 \ll m_e v_t^2$, which for heat flow in a temperature gradient is equivalent to
$k^{-1} \ll T/|\nabla T|$,
or (ii) $E_0 \ll v_t B_0$, which is equivalent to
the condition that the electron Larmor radius is much smaller than $T/|\nabla T|$.
These are also the conditions that allow us to assume 
that $F_0$ and $F_1$ are constant over a wavelength of the perturbation.

When discussing the features of the unstable modes, we will primarily focus on those relevant to thermal heat flow and non-thermal energy fluxes in dense plasmas
irradiated by a high power laser with relevance to IFE.

\section{The dispersion relation}

In the following derivations, the spatial differential operator $\nabla $ and the zeroth order magnetic and electric fields are aligned with the unit vector $\hat {{\bf x}}$
such that $\nabla \rightarrow ik\hat{\bf x}$.
First-order quantities oscillate such that $\partial {\bf \xi }_\perp /\partial t = - i \omega {\bf \xi }_\perp$
where
${\bf \xi }_\perp = \xi _y \hat {\bf y}+ \xi _z \hat{\bf z}$  is a general vector perpendicular to ${\bf B}_0$, giving
$$
-i \omega n_e m_A {\bf u}_\perp =n_e e {\bf E}_\perp- i  n_e e B_0 {\bf u}_\perp + {\bf R} _\perp 
\hskip 1  cm
\omega  {\bf B} _\perp = k \hat{\bf x} \times {\bf E} _\perp
\hskip 1 cm
ik \hat{\bf x} \times {\bf B} _\perp = \mu _0 {\bf J}_ \perp+\mu _0 n_e e {\bf u} _\perp \ .
\eqno{(12)}
$$
Circular polarisation is imposed such that 
$
\hat {\bf x} \times \xi _\perp = i \xi _\perp
$.
The sense (handedness) of polarisation is determined by the sign of $k$.
The terms dependent on the perturbed electron distribution in the VFP equation can be collected onto the left hand side of the equation to give
$$
 \frac {n-1}{2n-1} v \frac {\partial f_{n-1}^1}{\partial x}
+ \frac {n+2}{2n+3} v \frac {\partial f_{n+1}^1}{\partial x}
+ i \Omega _e f_n^1
+\frac {n(n+1) \nu (v) }{2} f_{n}^1
\hskip 6 cm
$$
$$
\hskip 3 cm 
=  \frac {\delta _{n1}}{2}
\bigg  \{ 
\frac {ie}{ m_e} (B_y - iB_z) F_1
+\frac {e}{m_e}  (E_y-iE_z)  \frac {\partial F_0}{\partial v}
- \nu (v)  (u_y-iu_z)\frac {\partial F_0}{\partial v}
\bigg \} \ .
\eqno{(13)}
$$
As shown in Appendix A for circular polarisation, equations (13) can be reconfigured to give
$$
{\bf f}_\perp (v) = g_1^* (v) \left \{
\frac {e {\bf B}_\perp}{m_e \Omega _e} F_1 
+\frac {ie {\bf E}_\perp}{m_e \Omega _e}\frac {\partial F_0}{\partial v} 
- i{\bf u}_\perp \frac {\nu}{\Omega _e}\frac {\partial F_0}{\partial v}
\right \}
\eqno{(14)}
$$
where
${\bf f}_\perp = 2\hat {\bf y} \Re (f_1^1)- 2 \hat {\bf z} \Im (f_1^1)$
is the vector part of the perturbed electron distribution function when expressed as a tensor expansion  (Johnston, 1960),
and $g_1(v)$ is the solution of the tridiagonal sequence of simultaneous equations $(n=1,2,3...)$
$$
\left ( 1 -  \frac {i n(n+1) \nu  }{2 \Omega _e }  \right ) g_n
+ \frac {n-1}{2n-1} \frac { k v}{\Omega _e}  g_{n-1}
+ \frac {n+2}{2n+3}  \frac { k v}{\Omega _e} g_{n+1}
= \delta _{n1} \ .
\eqno{(15)}
$$
The perturbed current and collisional force are given by 
$$
{\bf J}_\perp= -e \int _0 ^\infty \frac {4 \pi}{3} {\bf f}_\perp v^3 dv 
\hskip 1 cm
{\bf R}_\perp= m_e \int _0 ^\infty \frac {4 \pi}{3} {\bf f}_\perp v^3 \nu (v) dv 
+ {\bf u}_\perp  m_e \int _0 ^\infty \frac {4 \pi}{3} \frac {\partial F_0}{\partial v} v^3  \nu (v) dv
\ .
\eqno{(16)}
$$
As shown in Appendix B, these linearised equations can be combined to give the dispersion relation:
$$
(\omega - \omega _1)(\omega - \omega _2)+ \omega _3 (\omega - \omega _4)=0 \ .
\eqno{(17)}
$$
The frequencies $\omega _1$,  $\omega _2$,  $\omega _3$ and  $\omega _4$ are given by
$$
\omega _1 = \frac {k  I_{1g}}{I_{0g}}  -  \frac {k^2 c^2}{\omega _{pe}^2} \frac {\Omega _e}{I_{0g}} 
\hskip .9 cm
\omega _2= \Omega _i \left ( 1- i I_{0 \nu} - I_{0 g \nu \nu} \right )
\hskip .9 cm
\omega _3= \frac {\Omega _i (1 - iI_{0g \nu }  )^2 }{I_{0g}}
\hskip .9  cm
\omega _4 = \frac {kI_{1g \nu}}{I_{0g \nu} +i} 
\eqno{(18)}
$$
where $\Omega _i =(m_e/m_A) \Omega _e$ is the ion Larmor frequency,
and
$$  \hskip 2 cm
I_{0 \nu }= - \int _0^\infty \frac {4 \pi}{3 n_e} \frac {\nu}{\Omega _e} \frac {\partial F_0}{\partial v} v^3  dv
\hskip 2 cm
I_{0g}= - \int _0^\infty \frac {4 \pi}{3 n_e} \frac {\partial F_0}{\partial v} v^3 g_1 ^* (v) dv
\hskip 8 cm $$
$$  \hskip 2 cm
I_{0 g \nu}= - \int _0^\infty \frac {4 \pi}{3 n_e}  \frac {\nu}{\Omega _e} \frac {\partial F_0}{\partial v} v^3 g_1^* (v) dv
\hskip 1 cm
I_{0g \nu \nu}= - \int _0^\infty \frac {4 \pi}{3 n_e}  \frac {\nu ^2}{\Omega _e ^2} \frac {\partial F_0}{\partial v} v^3 g_1^* (v) dv
\hskip 8 cm
$$
$$ \hskip 2 cm
I_{1 g}= \int _0^\infty \frac {4 \pi}{3 n_e}  F_1 v^3 g_1^* (v) dv
\hskip 2.2 cm
I_{1 g \nu}=  \int _0^\infty \frac {4 \pi}{3 n_e}  \frac {\nu}{\Omega _e} F_1 v^3 g_1^* (v) dv \ .
\hskip 8 cm
\eqno{(19)}
$$
Equations (19) show that the dispersion relation depends on velocity integrals of $\partial F_0 /\partial v$ and $F_1$.
The integrals of $F_1$ drive the instability.
The integrals of  $\partial F_0 /\partial v$ represent the plasma response to the driving terms.
In many circumstances, the frequency $\omega _1$ and the integral $I_{1g}$ play the dominant role in driving the instability. 
In the case of stationary ions, $\Omega _i=0$
and the dispersion relation reduces to
$\omega = \omega _1$.

The function $g_1^*(v)$ describes the response of electrons at each velocity.
In a collisional plasma, the  instability is strong when $g_1^*(v)$ is real at some velocities but imaginary at others, thus representing different ways in which electrons
cross the magnetic field at different velocities.
In the next two sections we consider (i) thermal heat flow when the isotropic part of the electron distribution is Maxwellian, and (ii) an energy flux carried by a separate population of high energy streaming particles.

\section {A plasma carrying a heat flow}

In the Lorentz limit of high $Z$, and when non-local effects can be neglected, the zeroth order electron distribution function 
for Spitzer heat flow along a magnetic field is composed of isotropic and anisotropic parts
$$
F_0 (v) = \frac {n_e}{ v_t ^3 (2 \pi)^{3/2}} \exp(-v^2/2 v_t^2)
\hskip .5 cm {\rm and } \hskip .5 cm
F_1(v)=9.79\times 10^{-3} \frac{Q}{Q_{f}}
\left ( \frac {v^6}{v_t^6}-\frac {8v^4}{v_t^4} \right ) F_0 (v)
\eqno{(20)}
$$
respectively, where $Q/Q_f$ is the ratio of the heat flow $Q$ to the free-streaming heat flow $Q_f=n_em_e v_t^3$.
The expression for $F_1$ is derived by solution of of the 1D VFP equation either  for heat flow in the absence of a magnetic field or for heat flow aligned with a magnetic field
(Atzeni \& Meyer-ter-Vehn 2004, Craxton et al 2015).
The expression for $F_1$ can be related to the Braginskii thermal conductivity by substituting  $Q= - \kappa _\parallel \nabla T$
where $\kappa _\parallel = 13.58 v_t \tau Q_f/T$.

In ablating laser-produced plasmas, $Q/Q_f$ reaches a maximum of about 0.1.
As is well known (Gray \& Kilkenny 1980, Bell et al 1981), the heat flow is carried by a small fraction of low-collisionality electrons with velocities 
greater than $3 v_t$.
The majority of electrons are collisional and carry a return current to balance the forward current carried by the high velocity electrons.

There are four dimensionless free parameters in the system.
These can be chosen to be $Q/Q_f$, $ \Omega _e \tau$, $k v_t/\Omega_e$ \& $k^2 c^2 / \omega _{pe}^2$.
$Q/Q_f$ is the heat flow normalised to free-streaming, and represents the magnitude of the anisotropy.
The Hall parameter $ \Omega _e \tau$ is the ratio of the collision time to the Larmor gyration time.
$k v_t/\Omega_e$ is the ratio of the electron Larmor radius to the wavelength of the instability.

With four free parameters, a full parameter scan would require an extended discussion.
Instead, we plot the growth rate  against wavelength for a single standard reference case, relevant to laser-plasma experiments,
with the following parameters:
$n_e=10^{22}{\rm cm}^{-3}$,
$T= 1{\rm keV}$,
$B=1 {\rm MG}$,
$Q/Q_f=0.1$,
$Z \log \lambda =4$,
$A/Z=2$.
For these parameters, $\Omega _e=17.6 {\rm psec}^{-1}$, $\tau=0.27 {\rm psec}$, $c/\omega _{pe}=0.053 \mu{\rm m}$, $r_g=v_t/\Omega _e =0.75  \mu{\rm m}$,
giving $\Omega _e \tau = 4.8$.

Figure 3 plots the growth rate $\Im (\omega )$, where it is positive, against the wavelength $2 \pi /k$ for the standard parameters
for the two senses of polarisation.
In the left hand plot, the electrons carrying the heat flow rotate with the magnetic field as they propagate,
meaning that some of the heat-carrying electrons  track along a magnetic field line.
In this polarisation the instability is strongest when there are a large number of electrons moving 
resonantly with magnetic field lines. 
 This we label as the resonant polarisation.
Conversely we label the other
polarisation as non-resonant.
This designation derives from similarities with the resonant and non-resonant instabilities driven by streaming cosmic rays
(Bell 2014).
The right hand plot shows that the instability is present at long wavelengths in the non-resonant case in which
few electrons follow magnetic field lines.
The designation of polarisations as resonant and non-resonant is useful, but can be misleading since
the electrons carrying the return current rotate in the opposite direction in space and hence can be spatially resonant
when the heat-carrying electrons are non-resonant.

There are two roots to the quadratic dispersion relation, giving rise to two different modes that we label as an electron mode 
and an ion mode.  
The more slowly growing mode is designated as the ion mode because it disappears if
the ions are forced to be stationary.
With stationary ions, the dispersion relation is linear instead of quadratic,
and only the electron mode remains.
The curves in figure 3 for the growth rate of the electron mode are indistinguishably changed if the ions are forced 
to be stationary.

The resonant instability has the fastest growth rate.
With a growth time of about 10psec for these parameters, 
the resonant instability has ample time to grow during an IFE implosion,
potentially disrupting its symmetry.
However, the non-resonant instability may be more dangerous because it grows on a larger spatial scale while
still potentially growing through multiple e-foldings during an implosion.
The instability in either polarisation requires the pre-existence of a magnetic field that would be absent in
 an implosion with perfect spherical symmetry.
However, the symmetry can be broken by capsule imperfections or non-uniformities in laser energy deposition.
Pre-existing fields might be generated by a combination of the Biermann battery, non-local effects (Kingham \& Bell 2002),
resistive magnetic field generation (Davies et al 1997),
 the Rayleigh-Taylor instability, and other instabilities such as the Weibel instability.
A zeroth order magnetic field is inevitably present in experiments on planar targets irradiated 
by finite laser beams.

The instability in its resonant form is related to previously known instabilities driven by beams of electrons in collisionless plasmas, 
particularly occuring in the solar wind (see references in section 1).
It also has similarities with the unstable driving of Alfven waves by streaming cosmic rays in the interstellar medium (Lerche 1967, as reviewed by Wentzel 1974).
The non-resonant magneto-resistive instability at longer wavelengths has not to our knowledge been previously identified.
\begin{figure}
\includegraphics[angle=0,width=15cm]{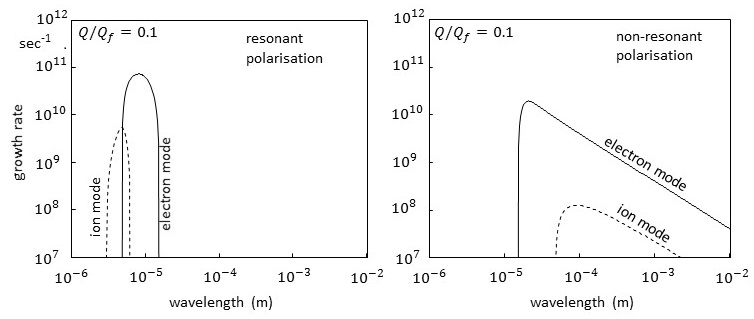}
\centering
\caption{
Log-log plot of the imaginary part of the frequency $\omega$
when the instability is driven by heat flow.
The left and right plots are for the resonant and non-resonant polarisations respectively.
The full lines are for the electron mode that 
is unchanged by forcing the ions to be stationary.
The dashed lines give the growth rate for the mode that is present only when ions are allowed to move.
}
\label{fig:figure3}
\end{figure}
\subsection {The limit of long wavelength}
An analytic expression for the growth rate can be derived in the limit in which ion motion is neglected, the wavelength is much larger than the algebraic mean of the electron Larmor radius and the electron mean free path,  and $Z\gg1$.  The neglect of ion motion simplifies the expression for the frequency to 
$\omega =kI_{1g}/I_{0g}$.

If additionally $\Omega _e \tau \gg 1$, then $g_1  ^* \rightarrow 1+i \nu/\Omega _e$, and
the frequency can be expressed as an expansion in powers of $1/(\Omega _e \tau)$.
The first term in the expansion is
imaginary,
$\omega = 0.138i  k v_t (Q/Q_f) (\Omega _e \tau)^{-1}$.  
The first real term in the expansion is proportional to $(\Omega _e \tau)^{-2}$.
Hence the mode is predominantly growing rather than advective in this limit.

If instead, $\Omega _e \tau \ll 1$, then $g_1  ^* \rightarrow -i \Omega _e /\nu + \Omega _e ^2/\nu ^2$, and
the frequency can be expressed as an expansion in powers of $(\Omega _e \tau)$.
The first two terms in the expansion are
$\omega = k  v_t (Q/Q_f)(0.72+10.18 i \Omega _e \tau)$.
In this collisional limit of small $\Omega _e \tau$, the dominant term is advective and the imaginary part of the frequency
is first order in $\Omega _e \tau$.

The growth rate in the limit of long wavelength is plotted in figure 6 over the full range of $\Omega _e \tau$.
The growth rate $\gamma _{Q,{\rm VFP}}$ is normalised to $kv_{Q0}$ where $v_{Q0}=v_t (Q/Q_f)$. 
The instability grows most rapidly when $\Omega _e \tau \sim 0.1-1$.
\section {A plasma carrying a current}
The previous section considered a plasma electron distribution function
consisting of an isotropic Maxwellian plus an anisotropic part carrying the heat flow and the return current.
In some laser-plasma experiments the energy deposited at low density by the laser is carried to high density by a separate, much more energetic, population of electrons,
while electrons at all velocities in the thermal population carry the return current.
The energetic electrons could be included in the above formalism by modifying $F_0$ and $F_1$ to include 
the high velocity population.
However, if energy-carrying electrons have energies much larger than the thermal electrons,
a simpler model can be adopted as formulated in section 2.
In this model, the high velocity electrons  are treated as a rigid uniform current 
that is undeflected by the magnetic field because of the much  larger Larmor radius of the electrons.

In this simpler model, 
the energetic population is removed from the analysis and replaced by the requirement that the thermal population
carries an imposed return electric current $j$.
The isotropic and anisotropic parts of the zeroth order distribution are then
$$
F_0 (v) = \frac {n_e}{ v_t ^3 (2 \pi)^{3/2}} \exp(-v^2/2 v_t^2)
\hskip .5 cm {\rm and} \hskip .5 cm
F_1(v)=0.0783 \frac{j}{j_{f}}
\frac {v^4}{v_t^4}  F_0 (v)
\eqno{(21)}
$$
where $j$ is parallel to the zeroth order magnetic field, and $j_f=n_e e v_t$ is the `free-streaming current'.
As for heat flow in section 5, the anisotropic part $F_1$ is derived by solution of the 1D VFP equation.
The expression for $F_1$ can be related to the electrical Braginskii resistivity by substituting  $j= E/\eta _\parallel$
where $\eta _\parallel = 0.294 m_e /n_e e^2 \tau$, which is equivalent to $j/j_f=3.40(e \tau /m_e v_t)E$.

When normalised to characteristic experimental values,
$$
\frac {j}{j_f}=0.12 \left ( \frac {Q_h}{10^{16}{\rm W  cm}^{-2}} \right )
 \left ( \frac {T_h}{100{\rm keV} } \right ) ^{-1}
 \left ( \frac {n_e}{10^{22}{\rm cm}^{-3}} \right )^{-1}
 \left ( \frac {T}{\rm keV} \right ) ^{-1/2}
\eqno{(22)}
$$
where $eT_h$ is the energy of the energy-carrying particles which are assumed to be monoenergetic.
$T$ is the temperature of the thermal electrons with a thermal velocity $v_t=(eT/m_e)^{1/2}$.
$Q_h$ is the energy flux carried by the energetic particles, which might be 0.1-30\% of the laser intensity,
depending on the efficiency of their generation and the geometry of their propagation.

\begin{figure}
\includegraphics[angle=0,width=15cm]{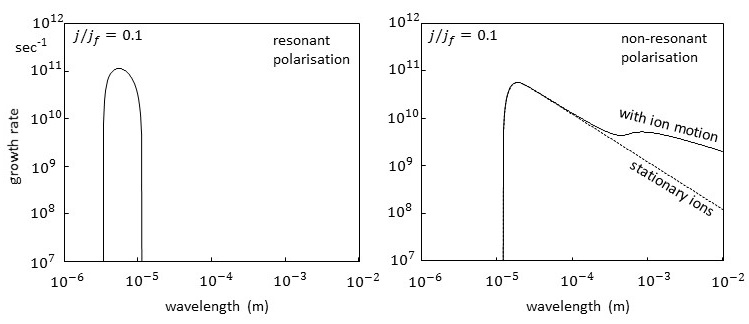}
\centering
\caption{
Log-log plot of the imaginary part of the frequency $\omega$
when the instability is driven by the presence of a return current balancing a rigid current carried by
streaming high energy charged particles.
The left and right plots are for the resonant and non-resonant polarisations respectively.
The full lines give the growth rate when ions are allowed to moved.
The dotted line gives the growth rate when the ions are forced to be stationary.
}
\label{fig:figure4}
\end{figure}

\begin{figure}
\includegraphics[angle=0,width=15cm]{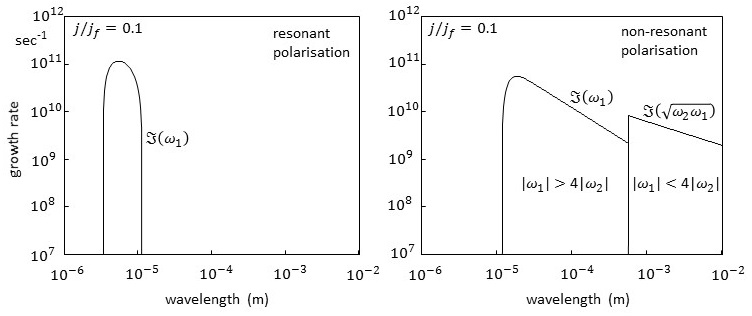}
\centering
\caption{
The imaginary parts of $\omega _1$ and $(\omega _1 \omega _2)^{1/2}$ for the two polarisations
when the instability is driven by the presence of a return current as in figure 4.
Comparison with figure 4 shows that these frequencies dominate the dispersion relation at small and large wavelengths respectively.
}
\label{fig:figure5}
\end{figure}

The growth rates are plotted in figure 4 for $j/j_f=0.1$.  The plasma parameters of density, temperature and magnetic field
are the same as those assumed in section 5 and figure 3.
From Braginskii (1965) the resistivities for these parameters are $\eta _\perp =1.1\times 10^{-6} {\rm Ohm\ m}$ and 
$\eta _\parallel =3.8\times 10^{-7} {\rm Ohm\ m}$.

For comparison with the illustrative calculation in section 2, these parameters give $\Delta \eta /\eta _0 =0.55$, $\Omega _e \tau = 4.8$ and
$v_t=1.33\times 10^7 {\rm m \ sec}^{-1}$.  
The growth rate given by equation 2 is then
$\gamma = 1.0\times 10^{12} {\rm sec} ^{-1} ({\rm wavelength}/{\mu {\rm m}})^{-1}$, which agrees with the line labelled `stationary ions' 
in the right hand plot of figure 4 for wavelengths exceeding $10 \mu {\rm m}$.
Resonance effects take over at wavelengths shorter than $10 \mu {\rm m}$.

The plots in figure 4 show features similar to those for the electron mode in figure 3.
However, there are differences.  
The dispersion relation is still quadratic (equations 17-19), but only one of the two modes is unstable.
The ion modes in figure 3 have disappeared from figure 4.
Nevertheless, ions still play a significant role.
Ion motion is responsible for the increased growth rate 
at long wavelengths at the right of the right-hand plot.
The dotted curve in figure 4 gives the growth rate when ion motion is neglected and the ions are forced to be stationary.
The full line shows that an increased growth rate is seen when ion motion is included.

The form of the curves in figure 4 for the growth rates can be understood by examining how the frequencies
$\omega _1$, $\omega _2$, $\omega _3$ \& $\omega _4$ affect the growth rate at different wavelengths.
$\omega _1$ and $\omega _4$ drive the instability since they consist of integrals of the anisotropy $F_1$.
$\omega _2$ \& $\omega _3$ contain the ion response since they are proportional to $\Omega _i$.
Another difference between $\omega _1$ \& $\omega _4$ and $\omega _2$ \& $\omega _3$ 
is that $ \omega _1$ \& $\omega _4$ are proportional to $k$ whereas  $\omega _2$ \& $\omega _3$ are independent
of $k$.
Hence, ion motion is relatively more important at long wavelengths.
This is responsible for the increased growth rate at large wavelengths in the right hand plot in figure 4
when ion motion is included.

Comparison of the growth rates plotted in figure 4 with $\omega _1$ and $(\omega _1 \omega _2)^{1/2}$ 
as plotted in figure 5 
shows that  $\omega =\omega _1$ is a good approximation at short wavelengths, but
ion motion takes over at long wavelengths where $\omega \approx (\omega _1 \omega _2)^{1/2} \propto k^{1/2}$.

If collisions are weak in the sense that $\Omega_e \tau \ll 1$, then 
$\omega _1 \rightarrow kj/n_e e$ and $\omega _2 \rightarrow \Omega _i$.
In this limit, the mode with $\omega = (\omega _1 \omega _2)^{1/2}$
transitions into the non-resonant mode responsible for the amplification of magnetic field (Bell 2004)
in diffusive shock acceleration of CR.

\subsection {The limit of long wavelength}
As with the instability driven by a heat flow,
the growth rate can be derived in the limit of large wavelength, $Z\gg 1$, no ion motion,
and $\Omega _ e \tau$ much greater or much less than 1.

If $\Omega _e \tau \gg 1$, $\omega = k  v_t  (j/j_f)(1-0.706i  (\Omega _e \tau)^{-1})$.  
In contrast to the heat flow-driven instability in section 5.1, the frequency has a real part to zeroth order in $\Omega _e \tau$.
This represents advection with the electron current.
The leading term in the imaginary part of the frequency
is proportional to $(\Omega _e \tau)^{-1}$
for both the current-driven and heat flow-driven instability.

If $\Omega _e \tau \ll 1$, $\omega = k  v_t (j/j_f) (1.932-7.312 i\Omega _e \tau)$.  
As with the heat flow-driven instabiity, the growth rate is proportional to $\Omega _e \tau$.

The growth rate in the limit of long wavelength is plotted in figure 6 over the full range of $\Omega _e \tau$.
The growth rate $\gamma _{j,{\rm VFP}}$ is normalised to $kv_{j0}$ where $v_{j0}=v_t (j/j_f)$. 
The instability grows most rapidly when $\Omega _e \tau \sim 0.1-1$.


\section{The instability as it appears in the Braginskii transport equations}

Heat flow, the return current and the electric field can be
described by the Braginskii equations (1965).
The Braginskii equations apply in the limit of small heat flow
or current and large spatial and temporal scales.
The Braginskii transport coefficients are derived by fitting polynomial approximations to solutions of the Boltzmann equation.
The starting point of Braginskii's analysis is equivalent to out starting point with the VFP equation.
Hence we should expect to find that a heat flow or current aligned with a magnetic field is unstable when described by the Braginskii equations.
We now show this to be the case by linearising the Braginskii equations for a circularly polarised perturbation.
 
The Braginskii equations include an Ohm's law relating  
the electric field ${\bf E}$ to an electron current ${\bf j}$, 
an electron pressure $P$,  and a temperature gradient $\nabla T$:
$$
-e n_e {\bf E} + {\bf j} \times {\bf B} - \nabla P
+ \alpha .{\bf j}/n_e e  -   \beta .\nabla T = 0
\eqno{(23)}
$$
where $\alpha $ and $\beta $  are tensors such that  
$$
\alpha .{\bf j} = \alpha _\parallel ({\bf b}.{\bf j}){\bf b}
+ \alpha _ \perp {\bf b} \times ({\bf j} \times {\bf b})
- \alpha _\wedge {\bf b} \times {\bf j} \hskip 1 cm 
$$
$$
\beta .\nabla T = \beta _\parallel ({\bf b}.\nabla T) {\bf b}
+ \beta _\perp {\bf b} \times (\nabla T \times {\bf b})
+ \beta _\wedge {\bf b} \times \nabla T
\eqno{(24)}
$$
where ${\bf b}={\bf B}/|{\bf B}|$.
$\alpha _\perp$ and  $\alpha _\parallel$ are related to  the resistivities $\eta _\perp$ and  $\eta _\parallel$ that appear above in section 2:
$\alpha _\perp =n_e^2 e^2 \eta _\perp$,  $\alpha _\parallel=n_e^2 e^2 \eta _\parallel$.
Alternative expressions for $\alpha $ and $\beta$ 
(with minor differences in notation) derived by Epperlein \& Haines (1986)
are in many circumstances more accurate than those of Braginskii,
but they can be less accurate when used to calculate to the differences $\alpha _\perp - \alpha _\parallel$
and  $\beta_\perp - \beta _\parallel$.

A temperature gradient has a scalelength  $L$ along the direction of the zeroth order heat flow such that $|\nabla T|=T/L$.
For the Braginskii equations to apply, $L$ must be larger than all other relevant scalelengths.

As in the VFP analysis, we couple these equations to the Maxwell equations,
but here for simplicity we assume that the ions are stationary.
The resulting equation for the perturbed magnetic field in a plasma with uniform density is
$$
\frac {\partial {\bf B}_\perp }{\partial t}=
\left [
\frac {(\alpha _\perp - \alpha _\parallel ){\bf B}_0.{\bf j} }{n_e^2 e^2 B_0^2} + \frac {( \beta _\parallel - \beta _\perp){\bf B}_0.\nabla T }{n_e eB_0 ^2} \right ] \nabla \times {\bf B}_\perp
+ \left ( \frac {\alpha _\perp}{n_e^2 e^2 \mu _0} \right ) \nabla ^2 {\bf B}_\perp
\hskip 8 cm 
$$
$$
+\left [ 
\left (1+  \frac {\alpha _\wedge }{n_e e B_0}  \right )  \frac {{\bf B}_0.{\bf j} }{n_e e B_0 ^2} +\frac { \beta _ \wedge {\bf B}_0.\nabla T }{n_e e B_0 ^3} 
\right ] ({\bf B}_0.\nabla ) {\bf B}_\perp
- \left ( 1  + \frac {\alpha _\wedge }{n_e e B_0 } \right ) \frac {({\bf B}_0 . \nabla )(\nabla \times {\bf B}_\perp)}{n_e e \mu _0}
\eqno{(25)}
$$
where ${\bf B}_0$ is the zeroth order magnetic field.
For a circularly polarised harmonic mode, as assumed in the VFP analysis in section 4,
this reduces to
$$
  \omega   =
i \left [
\frac {k ( \alpha _\parallel-\alpha _\perp  ){\bf B}_0.{\bf j}}{n_e^2 e^2 B_0 ^2}
- \frac {k ( \beta _\parallel - \beta _\perp){\bf B}_0.\nabla T}{n_e eB_0 ^2} \right ] 
-i\left [ \frac {\alpha_\perp}{n_e m_e} \frac {k^2 c^2}{\omega _{pe}^2} \right ]
\hskip 2 cm
$$
$$
\hskip 2 cm 
-  \left [ 
\left (1+ \frac {\alpha _\wedge }{n_e e B_0}  \right )  \frac {k{\bf B}_0.{\bf j} }{n_e e B_0 } 
+ \frac { k \beta _ \wedge {\bf B}_0.\nabla T }{ n_e e B_0^2} 
\right ] 
- \left [ \Omega _e
\left ( 1 + \frac {\alpha _\wedge }{n_e e B_0 } \right ) \frac {k^2 c^2}{\omega _{pe}^2} \right ]
 \ .
\eqno{(26)}
$$
The term in the first square bracket leads to stability or instability depending on the sign of $k$
and hence the sense of polarisation as in the VFP analysis.
The growth rate is proportional to the current $j$ or the temperature gradient $-\nabla T$ producing the heat flow.
The term in the second square bracket is the damping term that appears as  $k^2 c^2 /\omega _{pe}^2$ in the VFP analysis.
The term with the third square bracket advects the perturbed magnetic field with the current or the heat flow.
The term with the fourth square bracket is dispersive and  advective at a velocity proportional to $k$.

The first square bracket in equation 26 is the growth rate $\gamma $ which can be expressed 
as the sum 
$\gamma = \gamma _j + \gamma _Q$ of a growth rate $\gamma _j$ due to the current $\bf j$ and
a growth rate $\gamma _Q$ due to the heat flow $Q$.
Braginskii's polynomial approximations for $\alpha _\perp$, $\alpha _\parallel$, 
$\beta _\perp$ and $\beta _\parallel$  give
$$
\gamma _{j,{\rm Brag}} = kv_{0j}
\frac { (\alpha _1 ' \delta _0 - \delta _1 \alpha _0') \Omega _e \tau - \alpha _0 ' \Omega _e ^3 \tau^3}
{\delta _0 ( \delta _0 + \delta _1 \Omega _e ^2 \tau ^2 + \Omega _e ^4 \tau ^4)}
$$
$$
\gamma _{Q,{\rm Brag}} = kv_{0Q}
\frac { (\beta _1 ' \delta _0 - \delta _1 \beta _0') \Omega _e \tau - \beta _0 ' \Omega _e ^3 \tau^3}
{\gamma _0 \delta _0 ( \delta _0 + \delta _1 \Omega _e ^2 \tau ^2 + \Omega _e ^4 \tau ^4)}
\eqno{(27)}
$$
where $v_{0j}=v_t (j/j_f)$ and $v_{0Q}=v_t(Q/Q_f)$.
The numerical values of the constants $\delta$, $\alpha$ and $\beta$ can be found in Braginskii's table 2
for various values of $Z$.
For $Z=\infty$,
$$
\gamma _{j,{\rm Brag}} = - 6.75 kv_{0j} \ 
\frac { (1 + 1.088 \Omega _e ^2 \tau^2)  \Omega _e \tau}
{1+ 77.86  \Omega _e ^2 \tau ^2 + 10.41 \Omega _e ^4 \tau ^4}
$$
$$
\gamma _{Q,{\rm Brag}} = - 6.32 kv_{0Q}\ 
\frac { (1 + 0.201 \Omega _e ^2 \tau^2)  \Omega _e \tau}
{1+ 77.86  \Omega _e ^2 \tau ^2 + 10.41 \Omega _e ^4 \tau ^4} \ .
\eqno{(28)}
$$
\begin{figure}
\includegraphics[angle=0,width=10cm]{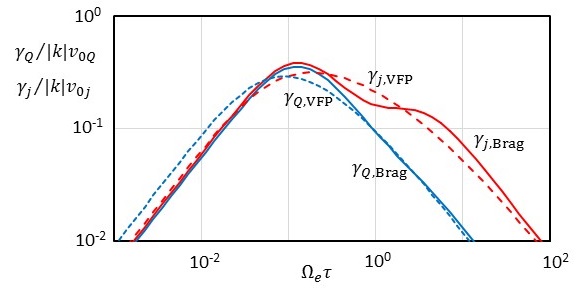}
\centering
\caption{
The Braginskii growth rates (equations 28) for the instability driven by a current (red full line)
and the heat flow (blue full line) normalised to $|k|v_{0j}$ and $|k|v_{0Q}$ respectively.
The dashed and dotted lines are the comparable growth rates when calculated from the VFP equation
in the limit of a large wavelength.
}
\label{fig:figure5}
\end{figure}
Instability occurs in the polarisation for which $k$ is negative.
These growth rates are plotted in non-dimensional form in figure 6 where they are compared with
the growth rates calculated from the VFP equation in the limit of long wavelength.
Figure 6 shows that the growth rate is largest when $\Omega _e \tau \sim 0.1 - 1$.

In the limit of small $\Omega _e \tau$,
$
\gamma _{j,{\rm Brag}} =-6.75 ( \Omega _e  \tau)  kv_{0j}
$
and
$
\gamma _{Q,{\rm Brag}} = - 6.32 ( \Omega _e  \tau)  kv_{0Q}
$.
In the limit of large $\Omega _e \tau$,
$
\gamma _{j,{\rm Brag}} =-0.705 ( \Omega _e  \tau) ^{-1}  kv_{0j}
$
and
$
\gamma _{Q,{\rm Brag}} = - 0.122 ( \Omega _e  \tau) ^{-1}  kv_{0Q}
$.

With allowance for Braginskii's method of approximation, the growth rates calculated from Braginskii agree with
the more accurate growth rates calculated directly from the VFP equation.
Braginskii's choice of polynomial approximation gives the correct dependence on
$\Omega _e \tau$ in the limits of small and large $\Omega_e \tau$. 
In contrast, the form of the polynomial approximations adopted by Epperlein \& Haines (1986) gives an incorrect dependence in the limit  $\Omega _ e \tau \ll 1$ in which the differences between 
$\alpha _\parallel$ \& $\alpha _\perp$ and $\beta _\parallel$ \& $\beta_\perp$
are very small.


\section{Conclusions }

We have solved the linearised electron VFP equations coupled to the Maxwell equations and the cold ion equation of motion 
for a magnetised collisional plasma.
The resulting dispersion relation contains a number of unstable modes.
Some of these modes are well known in their collisionless limit in other contexts as described in sections 1, 5 and 6.
Others are previously unknown to our knowledge.
We find that the Braginskii transport equations exhibit
the same instability in the appropriate limit.
The growth rates are large enough to make our analysis relevant to IFE experiments.
The instability driven by streaming cosmic rays during shock acceleration is contained as the collisionless limit within our dispersion relation.
Application to other plasmas is a subject for future analysis.

We have analysed only the linear phase of the instability.  
Its practical  importance will depend on its non-linear development and the amplitude at which it saturates.
The CR-driven form of the instability has been shown to grow by orders of magnitude beyond 
$\delta B/B_0 \sim 1$.
The existence of the non-linear non-resonant CR-driven instability has been confirmed by observation of supernova remnants (SNR) where initial fields of 
a few $\mu {\rm G}$ are seen to be amplified non-linearly to 100s $\mu {\rm G}$ (Vink \& Laming 2003, V\"{o}lk et al, 2005).
Substantial amplification is possible because the instability is non-resonant and is not hampered by  loss of resonance
when $\delta B $ exceeds $ B_0$.
Further theoretical analysis and experiment is needed to investigate the possibility of
magnetic field
amplification by large factors in the magneto-collisional case considered here.
If the magneto-collisional instability follows the behaviour of the CR-driven instability (eg Matthews et al 2017),
a fully non-linear numerical simulation will be needed.
A quasi-linear mode-coupling approach would not naturally capture the essential physics of this non-resonant instability.

One possible beneficial application of an amplified magnetic field might be to erect a transport barrier that shields the fuel from preheat
in IFE implosions, especially in the case of shock ignition where the laser intensity is raised at the end of the implosion to launch a shock into the fuel
(eg Betti et al 2007).

\section{Acknowledgements}
JHM acknowledges support from the Herchel Smith Postdoctoral Fellowship Fund.
Part of this work was funded through EPSRC doctoral training grant (No. EP/M507878/1).
We thank Alex Robinson (Rutherford Appleton Laboratory) for helpful discussions,
and we thank the referees for helpful comments.
 
\section{References}

Atzeni S, Meyer-ter-Vehn J 2009 \textit{The Physics of Inertial Fusion} publ Oxford University Press, USA.
\newline
Bell AR, Evans RG, Nicholas DJ 1981  \textit{Phys Rev Lett} \textbf{46} 243.
\newline
Bell AR 2004 \textit{MNRAS} \textbf{371} 550.
\newline
Bell AR 2014 \textit{Brazilian Journal of Physics} \textbf{44} 415.
\newline
Bell AR, Kingham RJ 2003 \textit{Phys Rev Lett} \textbf{91} 035003. 
\newline
Bell AR, Robinson APL, Sherlock M, Kingham RJ, Rozmus W 2006  \textit{Plasma Phys Cont Fusion} \textbf{48} 37.
\newline
Betti R, Zhou CD, Anderson KS, Perkins LJ, Theobald W, Solodov AA 2007 \textit{Phys Rev Lett} \textbf{98} 155001.
\newline
Bissell JJ, Ridgers CP, Kingham RJ (2010) \textit{Phys Rev Lett} \textbf{105} 175001.
\newline
Braginskii SI 1965 \textit{Rev Plasma Phys} \textbf{1}  205. 
\newline
Craxton RS plus 31 authors 2015  \textit{Phys Plasmas} \textbf{22} 22, 110501.
\newline
Davidson RC 1983  \textit{Handbook of Plasma Physics}, ed MN Rosenbluth \& RZ Sagdeev \textbf{1} 527. 
\newline
 Davies JR, Bell AR, Haines MG, Guérin SM 1997 \textit{Phy Rev E} \textbf{56} 7193.
\newline
Epperlein EM, Haines MG 1986 \textit{Phys Fluids} \textbf{29} 1029.
\newline
Forslund  DW 1970 \textit{J Geophys Res} \textbf{75} 17.  
\newline
Gary SP 1985 \textit{J Geophy Res} \textbf{90} 10815. 
\newline
Gary SP, Feldman WC, Forslund DW, Montgomery MD 1975  \textit{Geophys Res Lett} \textbf{2} 79. 
\newline
Gray DR, Kilkenny JD 1980 \textit{Plasma Phys} \textbf{22} 81.
\newline
Haines MG 1981 \textit{Phys Rev Lett} \textbf{47} 917.
\newline
Kingham RJ, Bell AR 2002 \textit{Phys Rev Lett} \textbf{88} 045004.
155001.
\newline
Lee S-Y, Lee E, Yoon HY 2019 \textit{ApJ} \textbf{876} 117. 
\newline
Lerche I 1967  \textit{ApJ} \textbf{147} L689. 
\newline
Lindl J  1995 \textit{Phys Plasmas} \textbf{2} 3933. 
\newline
Mackinnon AJ plus 26 authors 2004 \textit{Rev Sci Instr} \textbf{75} 3531.
\newline
Matthews JH, Bell AR, Blundell KM, Araudo AT 2017 \textit{MNRAS} \textbf{469} 1849
\newline
Miniati F, Bell AR 2011  \textit{ApJ} \textbf{729} 73. 
\newline
\textit{NRL Plasma Formulary (Washington DC: Naval Research Laboratory)} 2018 \textit{NRL/PU/6790--18-640}.
\newline
Rozmus W, Brantov AV, Sherlock M. Bychenkov VYu 2018 \textit{Plasma Phys Cont Fusion} \textbf{60} 014004.
\newline
Santos JJ plus 30 authors 2018 \textit{Phys Plasmas} \textbf{25} 056705. 
\newline
Schwartz SJ  1980 \textit{Rev Geo Space Sci} \textbf{18} 313.
\newline
Sherlock M, Bissell JJ 2020 \textit{Phys Rev Lett} \textbf{124} 055001.
\newline
 Snavely RA, Key, MH,  Hatchett SP, Cowan TE, Roth M,  Phillips  TW, Stoyer MA, Henry EA, Sangster TC, Singh MS, Wilks SC, MacKinnon A, Offenberger A, Pennington DM, Yasuike K, Langdon AB,  Lasinski BF, Johnson J, Perry MD, Campbell EM 2000 \textit{Phy Rev Lett} \textbf{89} 2945.
\newline
Spitzer L, Harm R 1953 \textit{Phys Rev} \textbf{89} 977.
\newline
Stamper JA, Papadopoulos K, Sudan RN, Dean SO, McLean EA, Dawson JM, 1971 \textit{Phys Rev Lett} \textbf{26} 1012.
\newline
Tidman DA, Shanny RA 1974 \textit{Phys Fluids} \textbf{17} 1207. 
\newline
Tzeferacos P plus 29 authors 2016 \textit{Nature comm} \textbf{9} 691.
\newline
Vink J, Laming JM, 2003 \textit{ApJ} \textbf{584} 758.
\newline 
V\"{o}lk HJ, Berezhko EG, Ksenofontov LT 2005 \textit{A\&A} \textbf{433} 229.
\newline
Weibel ES 1959 \textit{Phys Rev Lett} \textbf{2} 83. 
\newline
Wentzel DG 1974 \textit{ARA\&A} \textbf{12}  71.
\newline

\vskip .3 cm
\noindent {\bf Appendix A}

\noindent
Here we derive equation (14), as required for the dispersion relation.
Equation (13) expresses the perturbed distribution function in terms of the complex coefficients of equation (5).
We replace $\partial /\partial x$ with $ik$ to obtain
$$
 \frac {n-1}{2n-1} ikv f_{n-1}^1
+ \frac {n+2}{2n+3} ikv f_{n+1}^1
+ i \Omega _e f_n^1
+\frac {n(n+1) \nu  }{2} f_{n}^1
\hskip 6 cm
$$
$$
\hskip 3 cm 
=  \frac {\delta _{n1}}{2}
\bigg  \{ 
\frac {ie}{ m_e} (B_y - iB_z) F_1
+\frac {e}{m_e}  (E_y-iE_z)  \frac {\partial F_0}{\partial v}
- \nu   (u_y-iu_z)\frac {\partial F_0}{\partial v}
\bigg \}\ .
\eqno{(A1)}
$$
$f_1^1$ (the $n=1$ component of the sequence) is needed for the electron current and the resistive force (equations 16).
${\bf J }_\perp $ and ${\bf R} _\perp$ are integrals over $f_y$ and $f_z$ where these constitute the first order anisotropy in the tensor expansion
of the distribution function.
Johnston (1960) demonstrated the equivalence of the tensor and spherical harmonic expansions.
With our definition of the coefficients of the spherical harmonic expansion,
$f_y= 2 f_1^R$ and $f_z = - 2 f_1^I$  where $f_1^R = \Re (f_1^1)$ and $f_1^I= \Im (f_1^1)$.

We define the complex sequence of functions $g_n(v)$ such that
$$
f_n^1 (v)= - \frac {ig_n(v)}{2\Omega _e} \bigg  \{ 
\frac {ie}{ m_e} (B_y - iB_z) F_1
+\frac {e}{m_e}  (E_y-iE_z)  \frac {\partial F_0}{\partial v}
- \nu   (u_y-iu_z)\frac {\partial F_0}{\partial v}
\bigg \}
\eqno{(A2)}
$$
in which case, $g_1(v)$ can be found by solution of the tridiagonal  sequence of simultaneous equations,
$$
\left ( 1 -  \frac {i n(n+1) \nu  }{2 \Omega _e }  \right ) g_n
+ \frac {n-1}{2n-1} \frac { k v}{\Omega _e}  g_{n-1}
+ \frac {n+2}{2n+3}  \frac { k v}{\Omega _e} g_{n+1}
= \delta _{n1}\ .
\eqno{(A3)}
$$
The $n=1$ element of the set of equations $(A2)$ is 
$$
f_1^1= 
- 
\frac {g_1^R }{2\Omega _e} \bigg  \{ 
-\frac {e}{ m_e} (B_y - iB_z) F_1
+\frac {e}{m_e}  (iE_y+E_z)  \frac {\partial F_0}{\partial v}
- \nu   (iu_y+u_z)\frac {\partial F_0}{\partial v}
\bigg \}
\hskip 2 cm
$$
$$
+ 
\frac {g_1^I }{2\Omega _e} \bigg  \{ 
\frac {e}{ m_e} (iB_y + B_z) F_1
+\frac {e}{m_e}  (E_y-iE_z)  \frac {\partial F_0}{\partial v}
- \nu  (u_y-iu_z)\frac {\partial F_0}{\partial v}
\bigg \}
\eqno{(A4)}
$$
where $g_1= g_1 ^R +i g_1^I$, giving
$$
f_1^R= 
\frac {g_1^R }{2\Omega _e} \bigg  \{ 
\frac {e}{ m_e} B_y  F_1
- \frac {e}{m_e}  E_z  \frac {\partial F_0}{\partial v}
+\nu   u_z\frac {\partial F_0}{\partial v}
\bigg \}
+ 
\frac {g_1^I }{2\Omega _e} \bigg  \{ 
\frac {e}{ m_e} B_z F_1
+\frac {e}{m_e}  E_y  \frac {\partial F_0}{\partial v}
- \nu (v)  u_y\frac {\partial F_0}{\partial v}
\bigg \}
$$
$$
f_1^I= 
\frac {g_1^R }{2\Omega _e} \bigg  \{ 
-\frac {e}{ m_e}  B_z F_1
-\frac {e}{m_e}  E_y  \frac {\partial F_0}{\partial v}
+ \nu   u_y\frac {\partial F_0}{\partial v}
\bigg \}
+ 
\frac {g_1^I }{2\Omega _e} \bigg  \{ 
\frac {e}{ m_e} B_y  F_1
-\frac {e}{m_e}  E_z  \frac {\partial F_0}{\partial v}
+ \nu   u_z\frac {\partial F_0}{\partial v}
\bigg \} \ .
\eqno{(A5)}
$$
In the  tensor formalism, the vector electron drift is ${\bf f}_\perp = 2f_1^R \hat {\bf y}-2 f_1^I \hat{\bf z}$.
Similarly, we define 
${\bf B}_\perp = B_y \hat {\bf y}+ B_z \hat{\bf z}$, 
${\bf E}_\perp = E_y \hat {\bf y}+ E_z \hat{\bf z}$,
${\bf u}_\perp = u_y \hat {\bf y}+ u_z \hat{\bf z}$,
giving the purely real equation
$$
{\bf f}_\perp
= 
\frac {g_1^R}{\Omega _e} \left (
\frac {eF_1}{m_e} {\bf B}_\perp
+\frac {e}{m_e}  \frac {\partial F_0}{\partial v} \hat {\bf x} \times {\bf E}_\perp
- \nu \frac {\partial F_0}{\partial v} \hat {\bf x} \times {\bf u}_\perp
\right )
\hskip 2 cm
$$
$$
+
\frac {g_1^I}{\Omega _e} \left (
-\frac {eF_1}{m_e} \hat {\bf x} \times {\bf B}_\perp
+\frac {e}{m_e}  \frac {\partial F_0}{\partial v}  {\bf E}_\perp
- \nu \frac {\partial F_0}{\partial v}  {\bf u}_\perp
\right ) \ .
\eqno{(A6)}
$$
We now apply circular polarisation such that $\hat {\bf x} \times \rightarrow i$ ($\hat {\bf x} \times \xi _\perp = i \xi _\perp$), giving
$$
{\bf f}_\perp
= 
\frac {g_1^R}{\Omega _e} \left (
\frac {eF_1}{m_e} {\bf B}_\perp
+ i\frac {e}{m_e}  \frac {\partial F_0}{\partial v}  {\bf E}_\perp
- i \nu \frac {\partial F_0}{\partial v}  {\bf u}_\perp
\right )
+
\frac {g_1^I}{\Omega _e} \left (
- i \frac {eF_1}{m_e}  {\bf B}_\perp
+\frac {e}{m_e}  \frac {\partial F_0}{\partial v}  {\bf E}_\perp
- \nu \frac {\partial F_0}{\partial v}  {\bf u}_\perp
\right )
\eqno{(A7)}
$$
which is equivalent to
$$
{\bf f}_\perp
= 
\frac {g_1^*}{\Omega _e} \left (
\frac {eF_1}{m_e} {\bf B}_\perp
+ \frac {i e}{m_e}  \frac {\partial F_0}{\partial v}  {\bf E}_\perp
- i \nu \frac {\partial F_0}{\partial v}  {\bf u}_\perp
\right )
\eqno{(A8)}
$$
as required for equation (14) in the main text.

\vskip .3 cm
\noindent {\bf Appendix B}

\noindent
Here we derive the dispersion relation, equations (17-19).
When circular polarisation is imposed such that $\hat {\bf x} \times \xi _\perp = i \xi _\perp$,
equations (12) become
$$
-i \omega n_e m_A {\bf u}_\perp =n_e e {\bf E}_\perp- i  n_e e B_0 {\bf u}_\perp + {\bf R} _\perp 
\hskip 1  cm
\omega  {\bf B} _\perp = i k  {\bf E} _\perp
\hskip 1 cm
- k {\bf B} _\perp = \mu _0 {\bf J}_ \perp+\mu _0 n_e e {\bf u} _\perp
\eqno{(B1)}
$$
From equations (16) for ${\bf J}_\perp$ and ${\bf R}_\perp $,
$$
{\bf J}_\perp= -e \int _0 ^\infty \frac {4 \pi}{3} {\bf f}_\perp v^3 dv 
\hskip 1 cm
{\bf R}_\perp= m_e \int _0 ^\infty \frac {4 \pi}{3} {\bf f}_\perp v^3 \nu (v) dv 
+ {\bf u}_\perp  m_e \int _0 ^\infty \frac {4 \pi}{3} \frac {\partial F_0}{\partial v} v^3  \nu (v) dv
\eqno{(B3)}
$$
where from equation (14),
$$
{\bf f}_\perp = g_1^* \left \{
\frac {e {\bf B}_\perp}{m_e \Omega _e} F_1 
+\frac {ie {\bf E}_\perp}{m_e \Omega _e}\frac {\partial F_0}{\partial v} 
- i{\bf u}_\perp \frac {\nu}{\Omega _e}\frac {\partial F_0}{\partial v}
\right \} \ .
\eqno{(B3)}
$$
Together, these equations determine the dispersion relation.
Substituting the expression for ${\bf f}_\perp$ into the integrals for ${\bf J}_\perp$ and ${\bf R}_\perp$ gives 
$$
{\bf J}_\perp= 
-
\frac {e^2 {\bf B}_\perp}{m_e \Omega _e} 
\int _0 ^\infty \frac {4 \pi v^3 g_1^*}{3}  F_1 
dv
- 
\frac {ie^2 {\bf E}_\perp}{m_e \Omega _e}
\int _0 ^\infty \frac {4 \pi v^3 g_1^*}{3} \frac {\partial F_0}{\partial v} 
dv
+ 
 \frac {i e {\bf u}_\perp}{\Omega _e}
\int _0 ^\infty \frac {4 \pi v^3  \nu  g_1^*}{3} \frac {\partial F_0}{\partial v}
 dv
\hskip 3 cm 
$$
$$
{\bf R}_\perp= 
\frac {e {\bf B}_\perp}{ \Omega _e} 
\int _0 ^\infty \frac {4 \pi v^3 \nu g_1^*}{3}  F_1 
dv
+ 
\frac {ie {\bf E}_\perp}{\Omega _e}
\int _0 ^\infty \frac {4 \pi v^3 \nu g_1^*}{3} \frac {\partial F_0}{\partial v} 
dv
-
 \frac {i m_e {\bf u}_\perp}{\Omega _e}
\int _0 ^\infty \frac {4 \pi v^3  \nu ^2 g_1^*}{3} \frac {\partial F_0}{\partial v}
 dv
\hskip 3 cm
$$
$$
+ {\bf u}_\perp  m_e \int _0 ^\infty \frac {4 \pi v^3 \nu}{3} \frac {\partial F_0}{\partial v}  dv \ .
\eqno{(B4)}
$$
For ease of notation, equations (B4) can be written as 
$$
{\bf J}_\perp= 
-
\frac {n_e e^2 I_{1g} {\bf B}_\perp}{m_e \Omega _e} 
+
\frac {i n_e e^2 I_{0g} {\bf E}_\perp}{m_e \Omega _e} 
-
i n_e e I_{0g \nu} {\bf u}_\perp
\hskip 2.7 cm
$$
$$
{\bf R}_\perp= 
 n_e  e   I_{1g \nu} {\bf B}_\perp
-
i n_e  e I_{0g \nu}  {\bf E}_\perp
+
 i n_e m_e \Omega _e I_{0g \nu \nu } {\bf u}_\perp 
- 
 n_e m_e \Omega _e I_{0 \nu } {\bf u}_\perp
\eqno{(B5)}
$$
where
$$
I_{1g}=\int _0 ^\infty \frac {4 \pi v^3 g_1^*}{3 n_e}  F_1  dv
\hskip 1 cm 
I_{0g} = - \int _0 ^\infty \frac {4 \pi v^3 g_1^*}{3 n_e} \frac {\partial F_0}{\partial v}  dv
\hskip 1 cm 
I_{0g \nu}= - \int _0 ^\infty \frac {4 \pi v^3  \nu  g_1^*}{3 n_e \Omega _e } \frac {\partial F_0}{\partial v} dv
$$
$$
I_{1g \nu} = \int _0 ^\infty \frac {4 \pi v^3 \nu g_1^*}{3  n_e \Omega _e}  F_1 dv
\hskip 1 cm 
I_{0g \nu \nu }= - \int _0 ^\infty \frac {4 \pi v^3  \nu ^2 g_1^*}{3 n_e \Omega _e ^2 } \frac {\partial F_0}{\partial v} dv
\hskip 1 cm 
I_{0 \nu }= - \int _0 ^\infty \frac {4 \pi v^3 \nu}{3 n_e \Omega _e } \frac {\partial F_0}{\partial v}  dv \ .
\eqno{(B6)}
$$
Eliminating ${\bf J}_\perp$, ${\bf R}_\perp$ and ${\bf E}_\perp$  between equations (B1) and (B6) gives 
$$
\left [
\frac {kc^2 \Omega _e}{\omega _{pe}^2}
- I_{1g} 
+\frac {\omega I_{0g}}{k}
\right ]
{\bf B}_\perp
=
B_0
\Bigg [   i I_{0g \nu} -1   \Bigg ] {\bf u}_\perp
$$
$$
\Bigg [ 
\frac {\omega }{k}  (  iI_{0g \nu} -1 )
-  i I_{1g \nu} 
\Bigg ]
{\bf B}_\perp
=
B_0 
\Bigg [
1
- I_{0g \nu \nu } 
- i I_{0 \nu } 
- \frac {  \omega}{\Omega _i}
\Bigg ]
 {\bf u}_\perp
\eqno{(B7)}
$$
where $\Omega _i = (m_e/m_A) \Omega _e$.
These two equations can be combined to give the dispersion relation
$$
\left [
\frac {k^2 c^2 \Omega _e}{\omega _{pe}^2}
- k I_{1g} 
+ \omega I_{0g}
\right ]
\Bigg [
1
- I_{0g \nu \nu } 
- i I_{0 \nu } 
- \frac {  \omega}{\Omega _i}
\Bigg ]
-
\Bigg [ 
\omega  ( iI_{0g \nu}  - 1)  -  i k I_{1g \nu} 
\Bigg ]
\Bigg [    i I_{0g \nu} - 1  \Bigg ]
=0 \ .
\eqno{(B8)}
$$
We define the following frequencies:
$$
\omega _1 = \frac {k  I_{1g}}{I_{0g}}  -  \frac {k^2 c^2}{\omega _{pe}^2} \frac {\Omega _e}{I_{0g}} 
\hskip .8 cm
\omega _2= \Omega _i \left ( 1- i I_{0 \nu} - I_{0 g \nu \nu} \right )
\hskip .8 cm
\omega _3= \frac {\Omega _i (1 - iI_{0g \nu }  )^2 }{I_{0g}}
\hskip .8 cm
\omega _4 = \frac {kI_{1g \nu}}{I_{0g \nu} +i} \ .
\eqno{(B9)}
$$
The dispersion relation is then
$$
(\omega - \omega _1)(\omega - \omega _2)+ \omega _3 (\omega - \omega _4)=0
\eqno{(B10)}
$$
as required for equation (17) in the main text.

\end{document}